\documentclass{iopconfser}
\usepackage[switch]{lineno}
\usepackage{graphicx}
\usepackage[table]{xcolor}
\usepackage{makecell}
\usepackage{adjustbox}

\usepackage{cite}

\usepackage{hyperref}
\hypersetup{
  colorlinks = true,
  linkcolor  = blue,
  citecolor  = blue,
  urlcolor   = blue
}



\begin{document}
\title{3+1 GRHD simulations of NSBH mergers with light black holes using public codes.}

\author{S. Gomez Lopez$^{1,2}$, B. Giacomazzo$^{3,4}$ and F. Pannarale$^{1,2}$}

\affil{$^1$Dipartimento di Fisica, Sapienza Università di Roma, Roma, Italia}
\affil{$^2$Istituto Nazionale di Fisica Nucleare (INFN), Sezione di Roma, Roma, Italia}
\affil{$^3$Dipartimento di Fisica “Giuseppe Occhialini”, Università degli Studi di Milano-Bicocca, Milano, Italia}
\affil{$^4$Istituto Nazionale di Fisica Nucleare (INFN), Sezione di Milano-Bicocca, Milano, Italia}

\email{sebastian.gomezlopez@uniroma1.it}
\begin{abstract}
Recent observations of compact binary systems have provided evidence for black holes with masses below standard expectations. When paired with neutron stars (NSs), such low-mass black holes (BHs) have the potential to be detected by future multimessenger campaigns and could unveil unknown features of the BH population in the universe. Given the importance of accurate models for matched-filtering searches and Bayesian parameter estimation in discoveries of this sort, attention has renewed on the limitations of low-mass NSBH merger models and on the downstream impact of such limitations on gravitational-wave data analysis. Previous numerical relativity (NR) studies have revealed discrepancies in the predicted merger time, as well as more general mismatches and dephasing between phenomenological models and high-resolution simulations. The simulation presented here was motivated by these findings and illustrates how publicly available tools can be applied to this specific science case, towards informing next-generation, more refined gravitational-wave and kilonovae models. Here, we summarize the setup of a high-resolution general relativistic hydrodynamics simulation of an equal-mass NSBH system performed entirely with public codes (Einstein Toolkit and FUKA) and a finite-temperature equation of state. The results we show include the $(\ell,m)=(2,2)$ strain and 2D snapshots of the system undergoing tidal disruption and early postmerger disk formation. The system evolved for $\sim 4$ orbits while the $L_2$ norm of the Hamiltonian constraint remained below $6\times10^{-7}$ throughout the inspiral prior to growing near merger time.
\end{abstract}
\newpage

\section{Introduction}
\vspace{0.1cm}
With the first part of the Fourth Observing Run (O4a), the LIGO-Virgo-KAGRA Collaboration uncovered two more signals compatible with a neutron star--black hole (NSBH) coalescencing source \cite{LIGOScientific:2025slb}.  Interest in NSBHs has particularly intensified with the observation of the first two of these events, GW230529\_181500 \cite{gw230529}, and, independently, with the electromagnetic observations of PSR J0514--4002E \cite{PSR_J0514-4002E}, both providing evidence for unexpectedly low-mass black holes. These observational achievements have historically relied on increasingly accurate waveform models and large-scale data analysis pipelines built for matched-filtering searches and Bayesian parameter estimation on LIGO-Virgo-KAGRA data \cite{JaranowskiKrolakBook,Schutz-Sathya_2009}. Models describing the full inspiral--merger--ringdown (IMR) evolution are typically developed through a feedback loop between semi-analytical approaches and calibration to selected sets of numerical relativity (NR) simulations of the two-body problem in general relativity \cite{LEA,IMRPhenomNSBH,SEOBNRV4_ROM_NRTIDALV2_NSBH,NSBH_spin_HM}. For NSBH systems, however, only a handful of NR simulations exist in the equal-mass and near-equal-mass regime, most of them produced only recently with non-public codes \cite{Foucart_2019j, Markin_2023}, and current waveform models have not yet attempted to incorporate these limited datasets for calibration \cite{nsbh_review}.

This paper highlights the key technical aspects of the end-to-end implementation/configuration of a general-relativistic hydrodynamics (GRHD) simulation with public tools \cite{Matur_2024}, such as FUKA \cite{fuka_2021} and the Einstein Toolkit \cite{ETK_paper, ETK_zenodo}, and reports a few qualitative results from a single high-resolution GRHD simulation of an equal-mass NSBH merger with a finite-temperature equation of state. All results are reported in geometric units with $G=c=1$, where mass, length, and time share the same dimension. Solar masses are used as the reference across all plots and tables ($1\,M_\odot \approx 1.477\ \mathrm{km} \approx 4.93\times10^{-6}\ \mathrm{s}$), and quantities such as the rest-mass density and the Hamiltonian constraint are expressed consistently in these units.

\section{Initial data and Grid setup}
\vspace{0.1cm}

The high-resolution initial data were computed using \texttt{Fukav2} \cite{fuka_2021} on the CINECA Galileo100 cluster. The process took approximately 10 hours on 20 “Fat” nodes, each with 48 Intel Xeon Platinum 8276/L cores-per-node at 2.4 GHz, and reached a maximum RAM usage of about 275 GB-per-node. The NSBH solver was configured with 15 collocation points in the radial coordinate $r$ and the polar angle $\theta$, 14 points in the azimuthal angle $\phi$, eccentricity reduction enabled, and a solver accuracy of $10^{-8}$. 

\begin{figure}[b]
\centering
\begin{minipage}{0.60\textwidth}
\centering
\includegraphics[width=0.95\linewidth]{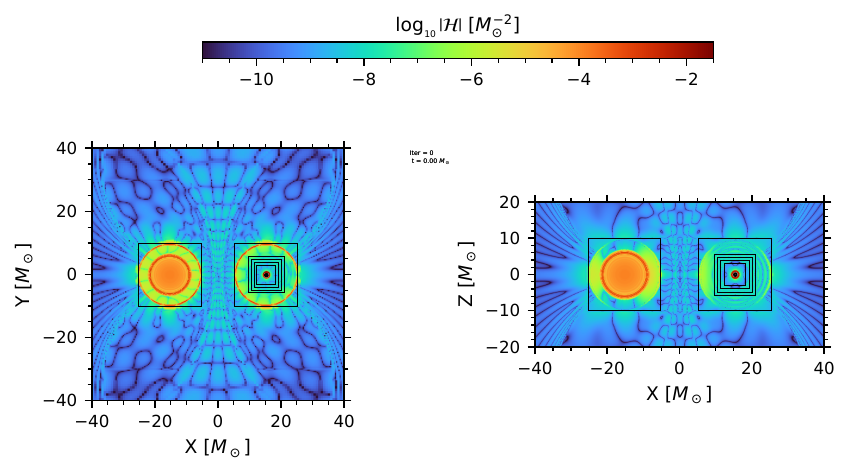}
\end{minipage}%
\hfill
\begin{minipage}{0.40\textwidth}
\centering
\scalebox{0.75}{
\begin{tabular}{|c c c c|}
\hline
\rowcolor{gray!30}
Ref. level & \makecell{Diameter \\ $[M_\odot]$} & \makecell{Resolution \\ $[M_\odot]$} & \makecell{Points \\ per dim.} \\ \hline
0 & 1920 & 24.0 & 80 \\
\rowcolor{gray!30}
1 & 360 & 12.0 & 30 \\
2 & 180 & 6.0 & 30 \\
\rowcolor{gray!30}
3 & 160 & 3.0 & 53 \\
4 & 100 & 1.5 & 66 \\
\rowcolor{gray!30}
5 & 80 & 0.75 & 106 \\
6 & 60 & 0.375 & 160 \\
\rowcolor{gray!30}
7 & 20 & 0.1875 & 106 \\
8 & 11.4 & 0.09375 & 121 \\
\rowcolor{gray!30}
9 & 9.6 & 0.046875 & 204 \\
10 & 7.8 & 0.0234375 & 332 \\
\rowcolor{gray!30}
11 & 6 & 0.01171875 & 512 \\ \hline
\end{tabular}
}
\end{minipage}
\caption{2D initial-data snapshots processed with \texttt{kuibit} \cite{Bozzola:2021hus}. Left: equatorial/meridional planes, coloured by log of the Hamiltonian constraint. Black squared contours mark the innermost-level refinement coverage on both objects. Right: table of adaptive mesh refinement levels.}\label{init_data}
\end{figure}

Two-dimensional snapshots of the Hamiltonian constraint illustrate \texttt{FUKA}’s use of \texttt{KADATH}’s infrastructure for large-scale parallel computing, i.e., spectral multi-domain decomposition and curvilinear grids~\cite{fuka_2021, kadath_2010,Markin_2023}. Figure~\ref {init_data} depicts how each compact object is covered by a central core domain surrounded by two spherical shells, while the region connecting them is spanned by multiple domains forming a bispherical grid.

\setlength{\tabcolsep}{4pt}

Within the Einstein Toolkit, the evolution grid is defined with \texttt{Carpet}’s adaptive mesh refinement \cite{Carpet}. The \texttt{FUKA} initial data are imported onto the adaptive mesh refinement hierarchy via \texttt{KadathThorn} and \texttt{KadathImporter}, which read the \texttt{.info}/\texttt{.dat} file pair and perform the interpolation. This pipeline populates \texttt{ADMBase} with the spatial metric $ \gamma_{ij} $ and extrinsic curvature $ K_{ij} $, and \texttt{HydroBase} with the rest-mass density $ \rho $, specific internal energy $ \epsilon $, and fluid three-velocity components $v^i $. Additionally, our simulation uses \texttt{KadathPizza} to process the electron fraction and temperature available in the tabulated Bombaci--Logoteta equation of state \cite{BL,BLhot}, and the lapse $ \alpha $ and shift $ \beta^i $ are reinitialized by the chosen gauge drivers after import.

We considered an equal-mass, non-spinning NSBH system with $1.4\,M_\odot$ component masses; at an initial separation of $30.47\,M_\odot$ ($45\,\mathrm{km}$), this system yielded a star with an $8.39\,M_\odot$ ($12.40\,\mathrm{km}$) areal radius and a black hole with initial areal radius --- as measured by \texttt{AHFinderDirect} \cite{AHFinderDirect} --- of $2.80\,M_\odot$ ($4.13\,\mathrm{km}$). Given the initial factor-of-three size difference between the two objects (see figure \ref{evol} top panel), we configured \texttt{Carpet} to cover the neutron star with $8$ refinement levels and the black hole with $11$ (see table and black contours in Fig.~\ref{init_data}).

\section{Evolution and physical observables}
\vspace{0.1cm}

\begin{figure}[t]
\centering

\includegraphics[width=0.95\linewidth]{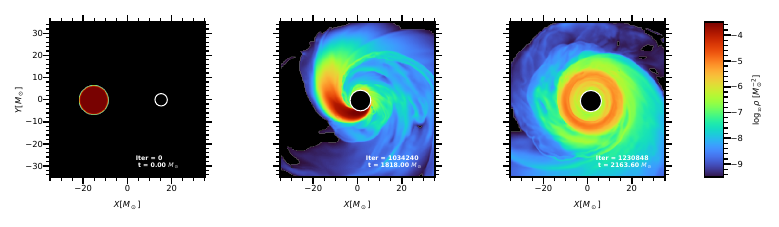}

\begin{minipage}{\linewidth}
  \centering
  \hspace*{-1.5cm} 
  \includegraphics[width=0.82\linewidth]{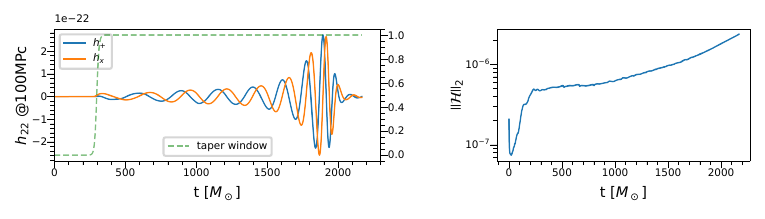}
\end{minipage}

\caption{Top: Rest-mass density snapshots (initial data, tidal disruption, disk formation); white contour shows BH's areal radius. Bottom: GW strain and Hamiltonian constraint ($L_2$ norm) evolution.}
\label{evol}
\end{figure}

To evolve spacetime, we imposed bitant symmetry across the $z=0$ plane and used the \texttt{Z4c} formulation of Einstein’s field equations \cite{Z4c} with damping parameters $\kappa_1=0.0$ and $\kappa_2=0.02$ to help reduce constraint violations \cite{BAM_nsbh}. During the $\sim 4$-orbit run, the $L_2$ Hamiltonian constraint remained below $6\times10^{-7}$ for the first $\sim 3$ orbits (see Fig.~\ref{evol} lower-right panel) before rising as tidal disruption sets in. We evolved spacetime and matter with \texttt{CTGamma} and  \texttt{WhiskyTHC}~\cite{THC,whiskythc1,wiskytch2} using $1+\log$ slicing and a Gamma-driver shift.

Gravitational radiation was extracted with the \texttt{WeylScal4} thorn, which computes the Newman--Penrose scalar $\Psi_4$ at several radii by projecting onto a null tetrad and decomposing into spin-weighted spherical harmonics using \texttt{Multipole} \cite{weyl_multipole_thorns,Ruiz_2007,GW_ext_bishop_rezzolla}. The GW strain rescaled to a detector at $100\,\mathrm{Mpc}$ was then obtained by double time integration of $\Psi_4$, using fixed-frequency integration to suppress low-frequency drifts \cite{psi4_int}. Figure~\ref{evol} (lower-left panel) shows the $(\ell,m)=(2,2)$ mode extracted at a coordinate radius of $200\,M_\odot$.


\section{Summary}
\vspace{0.1cm}
Systems like the one studied here are promising targets for multimessenger astronomy: they expose shortcomings in current gravitational-wave and kilonova models during the tidal disruption, merger, and post-merger phases. Improved modeling of such systems will help in achieving future detections, tightening constraints on the supranuclear-matter equation of state, and characterizing the black-hole population. To support follow-up work, we outlined key elements of our implementation on public infrastructure to aid reproducibility and community cross-checks, without attempting full documentation of all technical details.

\section{Acknowledgements}
\vspace{0.1cm}
The authors thank Samuel Tootle, Ivan Markin, and Tim Dietrich for valuable discussions during the course of this project. SGL thanks Rahime Matur and Margherita de Angelis for early contributions to parameter files. FP acknowledges support from the ICSC – Centro Nazionale di Ricerca in High Performance Computing, Big Data and Quantum Computing, funded by the European Union – NextGenerationEU, and from the Italian Ministry of University and Research (MUR) through the Progetti di ricerca di Rilevante Interesse Nazionale (PRIN) 2022 program (grant 20228TLHPE, CUP I53D23000630006). Numerical calculations were made possible by a CINECA–INFN agreement granting access to resources on LEONARDO at CINECA. High-resolution initial data were computed using a dedicated class-C allocation on Galileo100 at CINECA (HP10C5TI9T).

\bibliographystyle{iopart-num}
\bibliography{bib}
\end{document}